# 氢能或电能驱动人工光合作用固定 $CO_2$ 合成糖

**黄卫东**
中国科学技术大学地球与空间科学学院环境分部
合肥金寨路 96 号中科大七系 230026
email： huangwd@ustc.edu.cn tel: 86 551 3606631

**摘要：**植物光合作用是通过叶绿素吸收太阳能再生 NADH 和 ATP，驱动暗反应 Calvin－Benson 循环，合成糖。人工模拟光合作用，实现规模化合成糖的关键之一是低成本再生 ATP 与提高暗反应效率。本文提出 9 种人工光合作用暗反应途径，均只需要使用氢气或电再生 NADH 驱动，不需要再生 ATP 来驱动，而且提高了暗反应效率，与太阳能光伏或太阳能产氢技术结合，可以将粮食生产工业化，光合作用总效率可达 30%，比植物光合作用效率高一个数量级以上。其中利用氢气和二氧化碳化学合成甲醛驱动人工光合作用合成葡萄糖，仅需要 9 种酶，不需要不稳定的 NADH 和 ATP 参与，从而较容易实现规模化生产。该方法与胞外酶催化糖产氢方法一起，将使糖成为良好的储能载体。
关键词：人工光合作用，暗反应，二氧化碳固定，糖合成

# Synthesis of Sugar and fixation of $CO_2$ through Artificial Photosynthesis driving by Hydrogen or Electricity

Weidong Huang
Environmental Division, College of Earth and Space Science, University of Science and Technology of China, Hefei, China 230026
Abstracts: The overall process of photosynthesis consists of two main phases, the so-called light and dark eactions： light energy is absorbed by chlorophyll molecules and transferred to regenerate NADH and ATP, then drive Calvin-Benson cycle to synthesize sugar. In order to synthesize sugar through artificial photosynthesis, one of the key is to regenerate ATP economically and improve the efficiency of dark reactions. Here 9 kinds of dark reaction pathways are proposed, which only NADH is regenearated from hydrogen or electricity for driving, the efficiency of dark reactions is improved, combined with solar photovoltaic or solar hydrogen technology, the total efficiency of artificial photosynthesis can reach 30%, several ten times more than natural photosynthesis. One of them, to use chemical synthesis of formaldehyde from CO2 and H2, no NADH and ATP is need, to synthesize sugar efficiently through 9 enzymes, so it will be easier to produce in large scale, and the sugar will be a good energy carrier as the sugar can be efficiently converted to energy carrier hydrogen through enzymes.
Key words: artificial photosynthesis, dark reaction, CO2 fixation, synthesis of sugar

## 一 引言： 光合作用与人工光合作用

光合作用[1]是指植物和光合微生物吸收太阳能，将二氧化碳和水转化为碳水化合物的过程，它包括利用叶绿素吸收太阳能，并将能量转移到 NADPH 和 ATP 的光反应，和利用 NADPH 和 ATP 驱动，转化二氧化碳和水，形成碳水化合物的暗反应。由于化石燃料来自古代植物光合作用[2]，光合作用是当前人类食物和能量的主要来源，因而受到研究者的高度重视。

自然界光合作用效率较低，理论上，植物光合作用最大效率仅为 4.6－6.0%[3]，其主要原因包括[3]，第一，叶绿素只能吸收占太阳入射光能量不到一半的 400－700 纳米光子；第二，每吸收 8 个光子能量，只能产生转化一个分子二氧化碳所需要的 2 个 NADPH 分子和 3 个 ATP 分子，即使按照红光光子能量计算，也只有 43%的吸收能量转移到 NADPH 和 ATP 上，由于其他波长较短，能量较高，实际效率更低；第三，植物通过卡尔文循环，每消耗 12 摩尔 NADPH 和 18 摩尔 ATP，才能合成 1 摩尔葡萄糖，由于 1 摩尔葡萄糖氧化，释放的自由能仅为 686.8 千卡(高于燃烧释放的能量)，因此，按照自由能计算，卡尔文循环的热力学效率仅为 79%。发展人工光合作用，才能更好地提高效率。

广义上说，人工光合作用是指转化太阳能生产能源和各种化学品（如电和氢气）的方法[4]。人们

发展的太阳能光伏电池，将太阳能转化为电能，如多结砷化镓电池，效率达到 40%，未来预计可以达到 60%[5]，预计太阳能催化热解产氢的效率也大于 20%[6]，远远超过自然界植物光合作用的能量效率。

在人工光合作用合成糖方面，采用光伏电池或光伏产氢获取的能量再生 NADPH 和 ATP，驱动光合作用暗反应 Calvin 循环合成糖[4]，可提高光吸收及再生 NADPH 的能量效率，从而提高人工光合作用效率，如果实现大规模工业化生产粮食，与传统农业相比，可以大幅度降低土地需要量和水资源消耗。但是，光合作用合成 ATP，是通过光合磷酸化来进行的[1]，目前人工光合作用合成 ATP 的进展还很有限，最近的突破是第一次实现了人工光合作用产生 ATP 与 Calvin－Benson 循环的偶合[7]，但暗反应所消耗的 ATP 能量就接近产物糖的能量，考虑到植物 Calvin－Benson 循环中消耗的 ATP 所含有的能量仅相当于产物能量的 25%，其效率还很低。

我们此前根据胞外酶工程原理，组合多种生物体内代谢反应，提出一种人工光合作用合成糖的暗反应路线[8]，不需要消耗 ATP，从而不需要合成 ATP，但是，该过程需要 21 种酶和 NADPH 及 ATP 催化，才能合成等摩尔单糖和乙醇，能量转移到单糖的热力学效率仅为 61%；将酒精氧化，再生 NADPH，可提高糖转化效率到 82%，但是，这个人工光合作用合成糖的途径，共需要 34 种酶和 6 种辅酶，从而增加了工业化大规模生产的难度。

本文进一步分析综述了由酶催化代谢反应组成的人工光合作用固定二氧化碳，合成碳水化合物途径，只需要电或氢气再生 NADH，驱动合成糖，不需要消耗 ATP，其中最简单的途径，只需要 13 种酶，从而使人工光合作用实现难度显著降低。如果采用二氧化碳化学催化加氢合成甲醛，使用甲醛作为原料，只需要使用 9 种酶，不需要不稳定的 NADH 和 ATP，经过 11 步反应，就可以合成糖，从而容易实现。初步估算，我们提出的太阳能驱动人工光合作用固定二氧化碳合成糖的总效率，可达到 15－30%以上。

## 二 人工光合作用暗反应新途径

自然界光合作用暗反应，包括植物的 Calvin－Benson 循环，通过 Rubisco 酶固定二氧化碳，还原磷酸戊糖循环合成糖[1]；光合微生物发展了 5 种固定二氧化碳，合成乙酰辅酶 A 的路径[9]，再将乙酰辅酶 A 转化为糖。这些途径均由 NAD(P)H 和 ATP 共同驱动。本文提出的人工光合作用新途径，由 NADH 单独驱动，将水和二氧化碳合成为糖，如葡萄糖或淀粉，合成 1 摩尔葡萄糖只需要 12 摩尔 NADH，等于植物光合作用暗反应所需要的 NADH 数量，与植物光合作用同时还消耗 18 摩尔 ATP 相比，我们提出的人工光合作用途径，不需要消耗 ATP，NADH 可有电解再生或氢气还原再生，由电解再生 NADH 时，还与植物一样，同时产生氧气。我们提出的人工光合作用暗反应途径的总反应如下：

$2\,(C_6H_{10}O_5)_n + 12\,CO_2 + 12\,H_2O \rightarrow 12\,O_2 + 2\,(C_6H_{10}O_5)_{n+1}$ (电解)

$2\,(C_6H_{10}O_5)_n + 12\,CO_2 + 24\,H_2 \rightarrow 2(C_6H_{10}O_5)_{n+1}$ （氢气驱动）

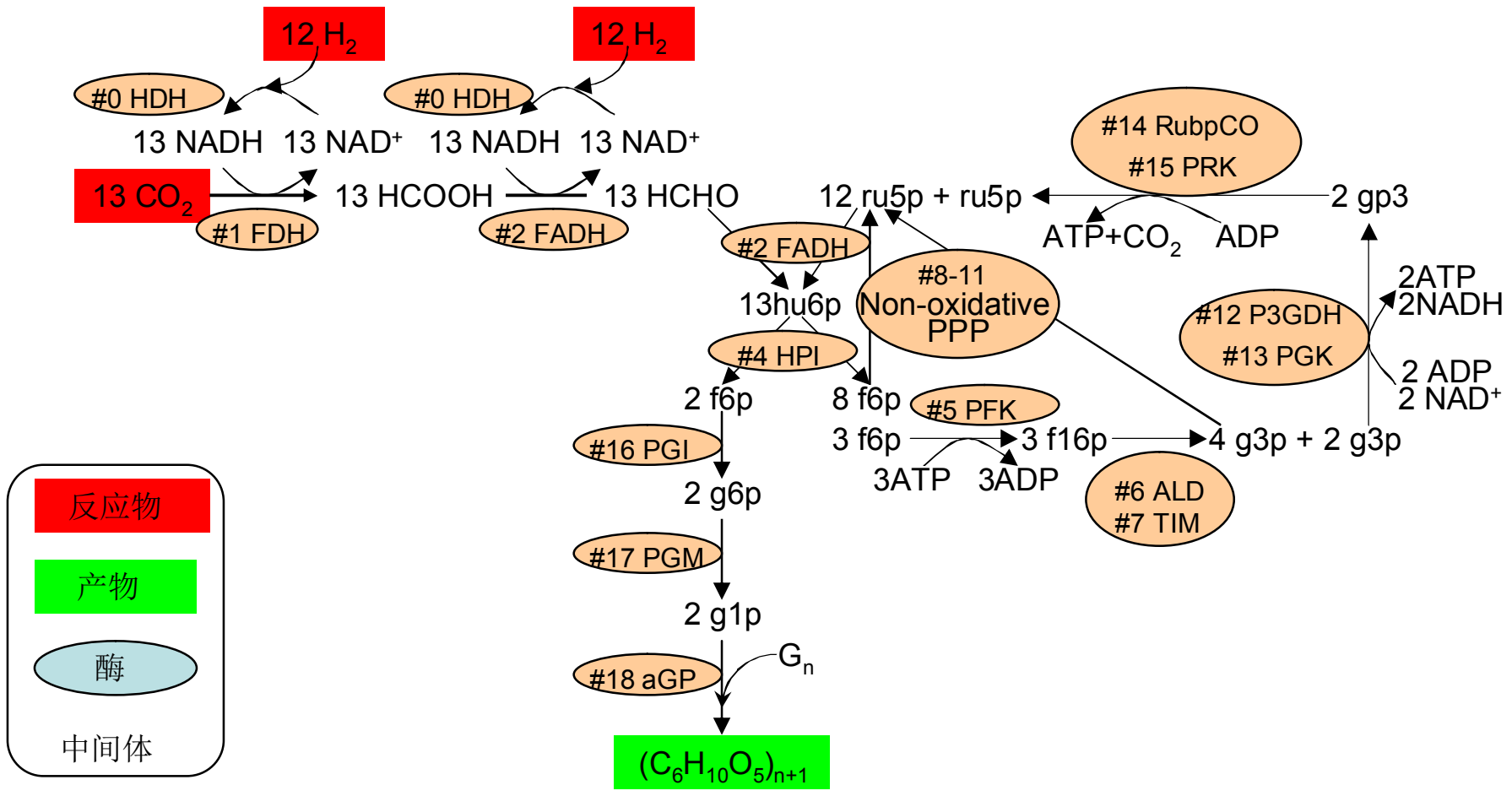

图 1. 电或氢能驱动固定二氧化碳人工光合作用途径：途径 1，磷酸甘油酸缩合途径
酶见表 1，中间产物：HCOOH，甲酸；HCHO，甲醛；ru5p，5 磷酸核酮糖；hu6p，6 磷酸 3 已酮糖；f6p，6 磷酸果糖；f16p，1，6 二磷酸果糖；g3p，3 磷酸甘油醛；gp3，3 磷酸甘油酸；g6p，6 磷酸葡萄糖；g1p，1 磷酸葡萄糖；Gn，直链淀粉；忽略了无机磷酸，$H^+$和水

图 1 是第一种氢气驱动人工光合作用途径示意图，总共使用了 19 种酶，催化了 20 步代谢反应（表 1），主要过程如下：

（1）氢气还原再生 NADH，使用 NAD 转氢酶催化，反应式如下：

**$24\ NAD^+ + 24\ H_2 \dashrightarrow 24\ NADH + 24\ H^+$**

如果使用电解还原 NAD，再生 24 摩尔 NADH，则反应式如下[10-12]：

**$24\ NAD^+ + 24\ H_2O \rightarrow 24\ NADH + 12\ O_2 + 24\ H^+$ (electrolysis)**

（2）由甲酸脱氢酶和甲醛脱氢酶催化，24 摩尔 NADH 驱动固定 12 摩尔二氧化碳，合成 12 摩尔甲醛[13-15]；此外，后续反应产生的 NADH 又固定 1 摩尔二氧化碳，合成 1 摩尔甲醛；总反应式如下：

$13\ CO_2 + 26\ NADH + 13\ H^+ \rightarrow 13\ HCHO + 26\ NAD^+$

（3）由磷酸己糖合成酶催化，将 13 摩尔甲醛缩合到 13 摩尔磷酸戊糖上，再通过异构酶转化为 13 摩尔磷酸果糖[16-17]。总反应式如下：

**13 HCHO + 13 ru5p → 13 f6p**

（4）由 8 摩尔磷酸果糖和 4 摩尔磷酸甘油醛通过磷酸戊糖途径再生合成 12 摩尔磷酸戊糖[1]；

**8 f6p + 4 g3p → 12 ru5p**

（5）使用 3 摩尔 ATP 驱动 3 摩尔磷酸果糖合成 3 摩尔二磷酸果糖，然后分解，再生 6 摩尔磷酸甘油醛[18]，其中 4 摩尔用于再生磷酸戊糖，另外 2 摩尔用于再生 ATP；

**3 f6p + 3 ATP → 6 g3p + 3 ADP**

（6）二摩尔磷酸甘油醛氧化，再生二摩尔 ATP 和 NADH，产生 2 摩尔磷酸甘油酸[19]；

$2\ g3p + 2\ Pi + 2\ ADP + 2\ NAD^+ \leftrightarrow 2\ gp3 + 2\ ATP + 2\ NADH + 2\ H^+$

（7）2 分子磷酸甘油酸偶联，再生一摩尔磷酸戊糖和二氧化碳，并再生另一摩尔 ATP[20-21]

$2\ gp3 + ADP \leftrightarrow Ru5P + ATP + H_2O + CO_2$

（8） 二摩尔磷酸果糖转化为磷酸葡萄糖，接续到淀粉链上[1]。

**2 f6p + 2 $(C_6H_{10}O_5)_n$ → 2 $(C_6H_{10}O_5)_{n+1}$ + 2 $P_i$**

表 1： 电或氢能驱动人工光合作用暗反应途径的反应和使用的酶：二磷酸甘油酸缩合途径

途径 1: 二磷酸甘油酸缩合途径

根据文献[51-52]数据计算自由能

| 序号 | 酶编号 | 酶名称 | 反应 | ΔrG'° kJ/m |
|---|---|---|---|---|
| **I. NADH 再生**<br>**1 电解: 24** $NAD^+$ + 24 $H_2O$ → 24 NADH + 12 $O_2$ + 24 $H^+$<br>**2 氢还原: 24** $NAD^+$ + 24 $H_2$ --> 24 NADH + 24 $H^+$ | | | | |
| | | | $NAD^+$ + $H_2O$ → NADH+$O_2$ + $H^+$ (电解) | 219.37 |
| #0 | 1.12.1.3 | 氢化酶 (HDH) | $NAD^+$ + $H_2$ --> NADH + $H^+$ | -17.82 |
| **II. 甲醛合成:** 13 $CO_2$ + 26 NADH + 13 $H^+$ → 13 HCHO + 26 $NAD^+$ | | | | |
| #1 | 1.2.1.2 | 甲酸脱氢酶 (FDH) | $CO_2$ + NADH → 甲酸 (HCOOH) + $NAD^+$ | **21.27** |
| #2 | 1.2.1.46 | 甲醛脱氢酶 (FADH) | 甲酸 + NADH + $2H^+$ → 甲醛 (HCHO) + $NAD^+$ + $H_2O$ | 41.00 |
| **III. 通过 5 磷酸核酮糖合成 6 磷酸果糖：13 $CH_2O$ + 13 ru5p → 13 f6p** | | | | |
| #3 | 4.1.2.43 | 6 磷酸已酮糖合成酶 (HPS) | 5 磷酸戊酮糖(ru5p) + 甲醛 → 6 磷酸已糖(hu6p) | -21.84 |
| #4 | 5.3.1.27 | 6 磷酸已酮糖变构酶 (HPI) | 6 磷酸已糖(hu6p) → 6 磷酸果糖 (f6p) | -7.32 |
| **IV. 5 磷酸核酮糖再生： 8 f6p + 4 g3p → 12 ru5p** | | | | |
| #5 | 2.2.1.1 | 转酮酶 (TK) | 6 磷酸果糖 (f6p) + 3 磷酸甘油醛 (g3p) → 5 磷酸木酮糖 (x5p) + 4 磷酸赤藓糖 (e4p)<br>7 磷酸景天庚酮糖(s7p) + 3 磷酸甘油醛 (g3p)→ 5 磷酸木酮糖 (xu5p) + 5 磷酸核糖 (r5p) | 7.32<br>-8.70 |
| #6 | 2.2.1.2 | 醛酸移转酶 (TAL) | 6 磷酸果糖(f6p)+ 4 磷酸赤藓糖(e4p)→ 7 磷酸景天庚酮糖(s7p) + 3 磷酸甘油醛(g3p) | 7.32 |
| #7 | 5.3.1.6 | 5 磷酸核糖异构酶 (R5PI) | 5 磷酸核糖(r5p) → 5 磷酸核酮糖(ru5p) | 2.59 |
| #8 | 5.1.3.1 | 5 磷酸核酮糖差向酶 (Ru5PE) | 5 磷酸木酮糖(x5p) →5 磷酸核酮糖(ru5p) | 0 |
| **V. 再生用于合成 5 磷酸核酮糖的 3 磷酸甘油醛： 3 f6p + 3 ATP → 6 g3p + 3 ADP** | | | | |
| #9 | 2.7.1.11 | 6 磷酸果糖激酶(PFK) | 6 磷酸果糖 (f6p) + ATP → 1，6 二磷酸果糖 (f16p) +ADP | -18.83 |
| #10 | 4.1.2.13 | 二磷酸果糖酶 (ALD) | 1，6 二磷酸果糖 (f16p) → 3 磷酸甘油醛 (g3p) + 磷酸二羟丙酮 (dhap) | 18.03 |
| #11 | 5.3.1.1 | 磷酸丙糖异构酶 (TIM) | 磷酸二羟丙酮(dhap) → 3 磷酸甘油醛 (g3p) | 6.11 |
| **VI. 将 3 磷酸甘油醛转化为 3 磷酸甘油酸，同时再生 1 摩尔 ATP:**<br>**2g3p+2Pi+2ADP+2$NAD^+$ ↔ 2gp3+ 2ATP +2NADH + $2H^+$** | | | | |
| #12 | 1.2.1.12 | 3 磷酸甘油醛脱氢酶 (G3PDH) | 3 磷酸甘油醛 (g3p) + $NAD^+$ + Pi→1，3 二磷酸甘油酸 (bpg)+ NADH | -1.51 |
| 13 | 2.7.2.3 | 3 磷酸甘油酸激酶 (PGK) | 1，3 二磷酸甘油酸(bpg) + ADP → 3 磷酸甘油酸 (gp3) + ATP+$H^+$ | -11.0 |
| **VII. 3 磷酸甘油酸转化为 5 磷酸核酮糖，再生 ATP：2gp3+ADP←→ Ru5P + ATP + $H_2O$ + $CO_2$** | | | | |
| 14 | 4.1.1.39 | 二磷酸核酮糖羧化酶 (RubpCO) | 23 磷酸甘油酸(gp3) + 2 $H^+$ →1，5 二磷酸核酮糖 (rubp) + H2O + $CO_2$ | 26.02 |
| 15 | 2.7.1.19 | 磷酸核酮糖激酶 (PRK) | ADP + 1，5 二磷酸核酮糖(rubp)→ ATP +D5 磷酸核酮糖(ru5p) | 17.55 |
| **VIII. 由 6 磷酸果糖合成糖： 2f6p + $2G_n$ →2 $G_{n+1}$ + $2H^+$ + $2P_i$** | | | | |
| #16 | 5.3.1.9 | 磷酸葡萄糖异构酶 (PGI) | 6 磷酸果糖 (f6p) → 6 磷酸葡萄糖 (g6p) | 3.72 |
| #17 | 5.4.2.2 | 葡萄糖磷酸变位酶 (PGM) | 6 磷酸葡萄糖 (g6p) →1 磷酸葡萄糖 (g1p) | 0 |
| #18 | 2.4.1.1 | 淀粉磷酸化酶 (αGP) | $(C_6H_{10}O_5)_n$ ($G_n$) + 1 磷酸葡萄糖 (g1p) → $(C_6H_{10}O_5)_{n+1}$ ($G_{n+1}$) + $P_i$ +H+ | 5.54 |
| **总反应：** | | | 2 $(C_6H_{10}O_5)_n$ + 12 $CO_2$ + 12 $H_2O$ →12$O_2$ + 2$(C_6H_{10}O_5)_{n+1}$ (electricity)<br>2 $(C_6H_{10}O_5)_n$ + 12 $CO_2$ + 24 $H_2$ →2$(C_6H_{10}O_5)_{n+1}$ | 6345.5<br>21.48 |

总共 19 种酶，催化 20 个反应

此前提出的路径[8]是由磷酸甘油醛分解再生 ATP，副产乙醇，这使四分之一原料转化为酒精。虽然乙醇可进一步通过三羧酸循环分解再生 NADH，但是，这需要增加 13 种酶和 4 种辅酶，使系统吸引 34 种酶，6 种辅酶，变得很复杂，尤其中间产物增多。本方法是将 2 摩尔磷酸甘油醛水解氧化产生磷酸甘油酸，再生 2 摩尔 ATP，然后一步缩合磷酸甘油酸再生 1 摩尔 ATP 和磷酸戊糖（简称磷酸甘油酸缩合途径），从而减少了酶和中间产物数量，使需要的酶数量减少到 19 种。

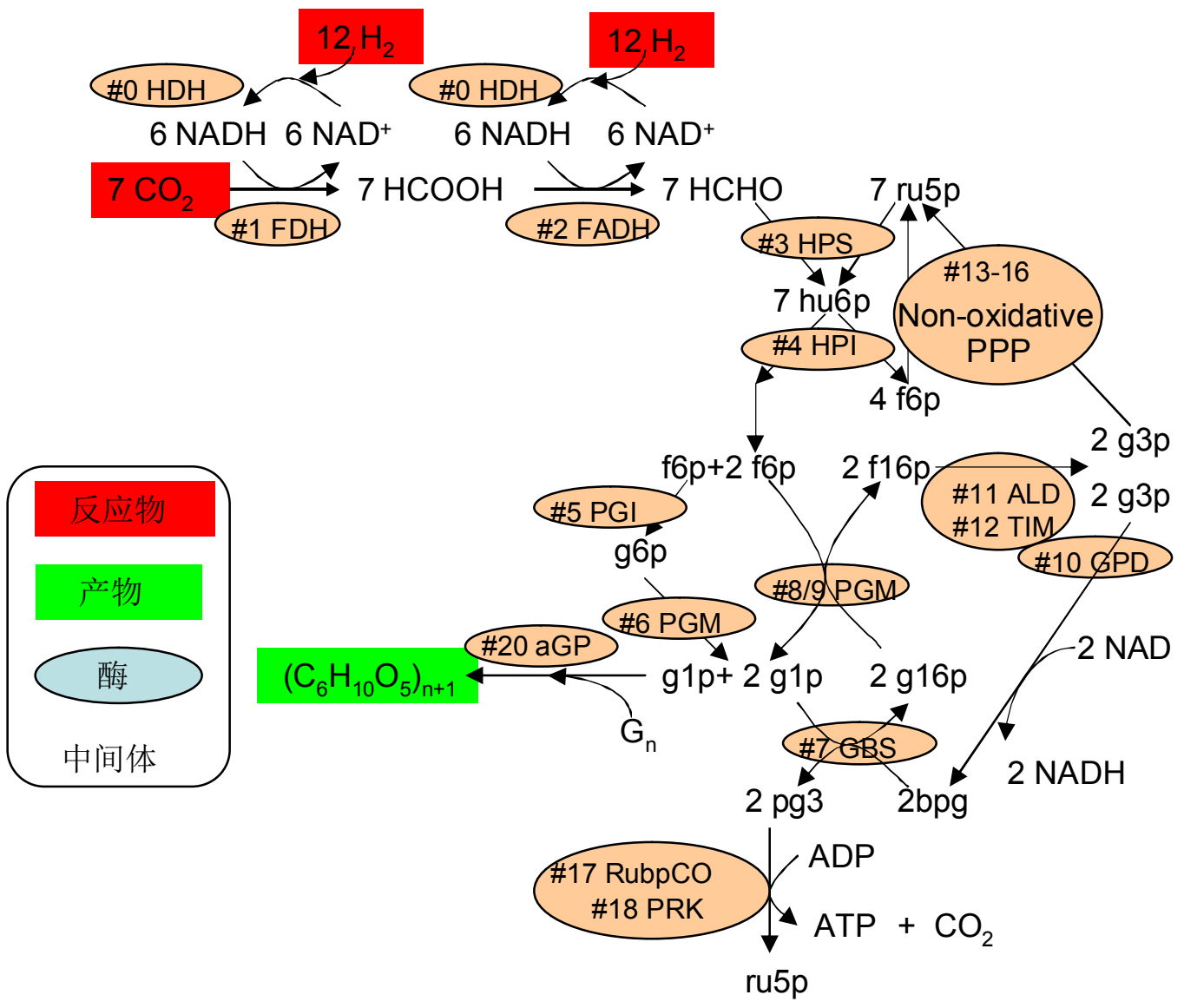

图 2. 电或氢能驱动固定二氧化碳人工光合作用途径：途径 2，二磷酸甘油酸合成途径

与二磷酸甘油酸缩合途径不同的酶见表 2，中间产物不同图 1 的如下：g16p，1，6 二磷酸葡萄糖；ga6p，6 磷酸半乳糖；省略无机磷酸，$H^+$和水。

由于上述方案需要 ATP 和各种磷酸脂参与合成糖，它们容易水解，从而消耗起催化作用的中间体，增加了合成糖的能耗和成本，因此，我们提出了一个能够增殖 ATP，从而补充各种中间体的新路径，参见图 2，与前述途径相比，改变了再生二磷酸果糖途径，主要反应如下：

1、 通过二磷酸葡萄糖合成酶催化 1 磷酸葡萄糖与二磷酸甘油酸作用，产生二磷酸葡萄糖（g16p） 和磷酸甘油酸[22]：
g1p + bpg → g16p + gp3 + $H^+$

2、 二磷酸葡萄糖再通过磷酸葡萄糖变位酶催化，与磷酸果糖反应，再生二磷酸果糖[23]。副产物 galactose-6-phosphate 在该酶催化下，转变为磷酸葡萄糖：
f6p + g16p →f16p + g6p

其他过程均与前面途径相同，这个循环过程是通过二磷酸甘油酸合成二磷酸果糖，简称为**二磷酸甘油酸合成途径**，产糖主反应不需要消耗 ATP 来合成二磷酸果糖，副产物 ATP 可用于磷酸化，补充磷酸脂类中间产物水解带来的损耗，从而维持系统运行，又可称为 **ATP 增殖途径**。该方案需要 19 种酶，催化 22 种反应，包括一个增加二磷酸果糖酶合成补充二磷酸果糖（表 2 中反应#20）反应，由于各种中间产物磷酸糖脂是相互循环再生的，只要再生一种中间体，如二磷酸果糖，就可以弥补各种磷酸脂水解带来的影响。

表 2： 电或氢能驱动人工光合作用暗反应途径的反应和使用的酶：二磷酸甘油酸合成途径
使用下列 3 个反应代替途径 1 中#9 和#13 两个分那样，增加#4 反应，用于再生磷酸化中间体，补充水解损失，根据文献[51-52]数据计算自由能

| 序号 | 酶编号 | 酶名称 | 反应 | ΔrG'° kJ/m |
|---|---|---|---|---|
| #1 | 2.7.1.106 | 1，6 二磷酸葡萄糖合成酶（GBS) | 1 磷酸葡萄糖（g1p）+ 1,3 二磷酸甘油酸(bpg) → 1,6 二磷酸葡萄糖(g16p) + 3 磷酸甘油酸(gp3) + $H^+$ | 6.32 |
| #2 | 5.4.2.2 | 磷酸葡萄糖变位酶 (PGM) | 6 磷酸果糖（f6p) + 1,6 二磷酸葡萄糖(g16p) → 1,6 二磷酸果糖(f16p) +6 磷酸半乳糖(ga6p) | 0 |
| #3 | 5.4.2.2 | 磷酸葡萄糖变位酶 (PGM) | 6 磷酸半乳糖(ga6p) <==>1 磷酸葡萄糖（g1p) | 0 |
| | | | | |
| #4 | 2.7.1.11 | 6-磷酸果糖激酶（PFK) | 6 磷酸果糖（f6p）+ ATP → 1,6 二磷酸果糖(f16p) +ADP | -18.83 |

* 总共 19 个酶，催化 22 个反应，其中磷酸葡萄糖变位酶(5.4.2.2)催化了 3 个反应，转酮酶(2.2.1.1)催化了两个反应。

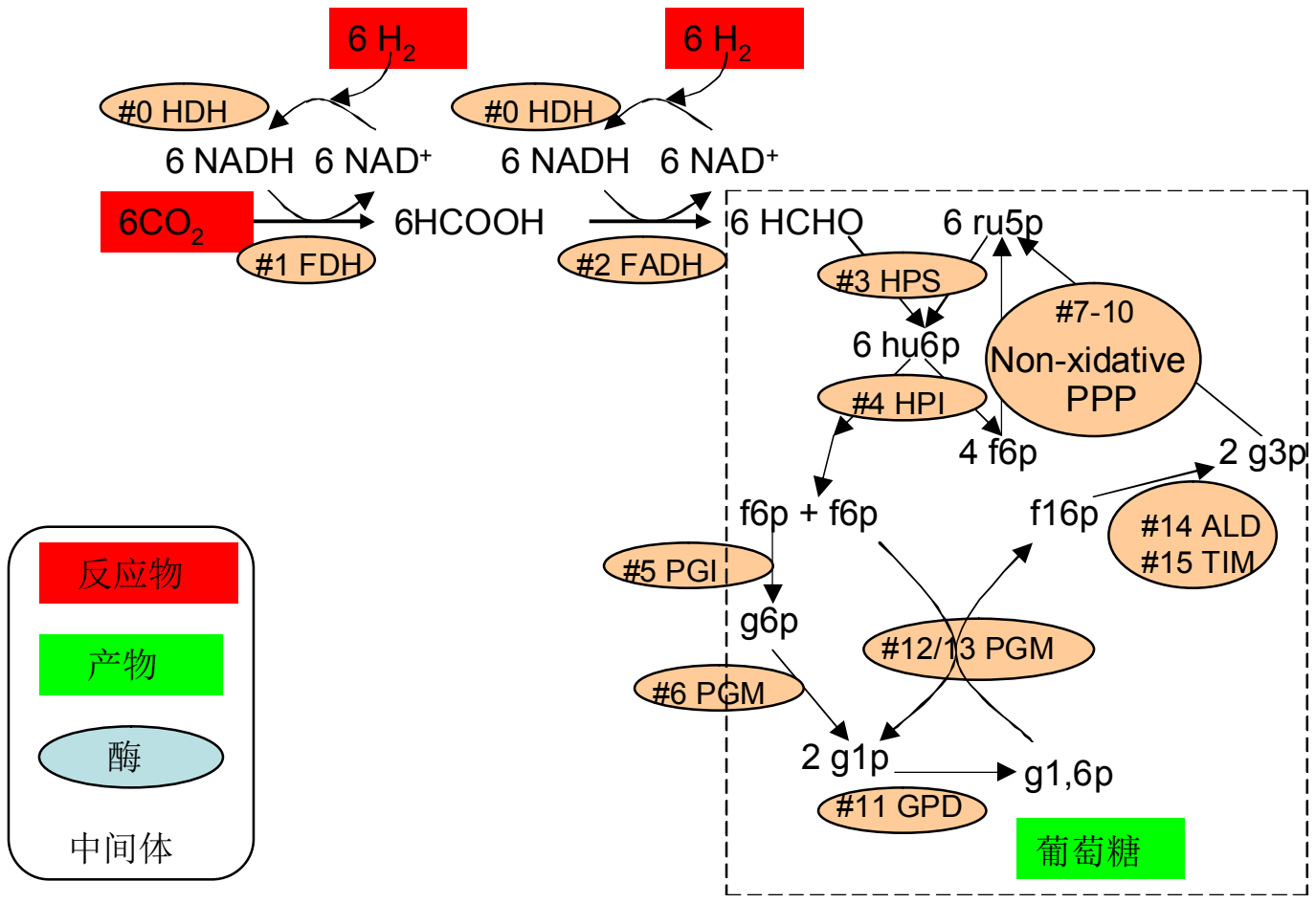

图 3. 电或氢能驱动固定二氧化碳人工光合作用途径：磷酸葡萄糖歧化途径
与二磷酸甘油酸缩合途径不同的酶见表 3，中间产物不同图 1 和 2 的如下；glucose，葡萄糖；省略无机磷酸，$H^+$和水。

我们提出的第三种转化途径，过程更简单（图 3），同样改变二磷酸果糖再生过程，产物是葡萄糖，其反应如下：

1、 2 分子 1 磷酸葡萄糖在歧化酶作用下，产生葡萄糖和二磷酸葡萄糖[22]：

2 g1p → glucose + g16p

2、 二磷酸葡萄糖再通过磷酸葡萄糖变位酶催化，与磷酸果糖反应，再生二磷酸果糖[23]：

f6p + g16p → f16p + g6p

这个途径是通过二磷酸葡萄糖歧化再生二磷酸果糖，简称为**磷酸葡萄糖歧化途径**，整个过程，不需要 ATP 参与，从而消除了 ATP 水解带来的影响。总共只需要 14 种酶和 NADH 参与催化（表 3），从而进一步简化了人工光合作用暗反应途径。

表 3： 电或氢能驱动人工光合作用暗反应途径的反应和使用的酶，途径 3：磷酸葡萄糖歧化途径
在途径 1 中使用下列反应代替反应#9，#12－15，#18，途径 1 中反应#16 将进行逆反应。根据文献[51-52]数据计算自由能

| 序号 | 酶编号 | 酶名称 | 反应 | ΔrG'° kJ/mol |
|---|---|---|---|---|
| #11 | 2.7.1.41 | 1 磷酸葡萄糖歧化酶 (GPD) | 2 葡萄糖 1 磷酸(g1p) → 1,6 二磷酸葡萄糖 (g16p) + 葡萄糖 | -2.62 |
| #12 | 5.4.2.2 | 磷酸葡萄糖变位酶 (PGM) | 6 磷酸果糖 (f6p) + 1,6 二磷酸葡萄糖 (g16p) → 1,6 二磷酸果糖 (f16p) + 6 磷酸半乳糖 (ga6p) | 0 |
| #13 | 5.4.2.2 | 磷酸葡萄糖变位酶 (PGM) | 6 磷酸半乳糖 (ga6p) →1 磷酸葡萄糖 (g1p) | 0 |

* 总共 14 种酶，催化 17 种反应，如果由电驱动，仅需要 13 种酶

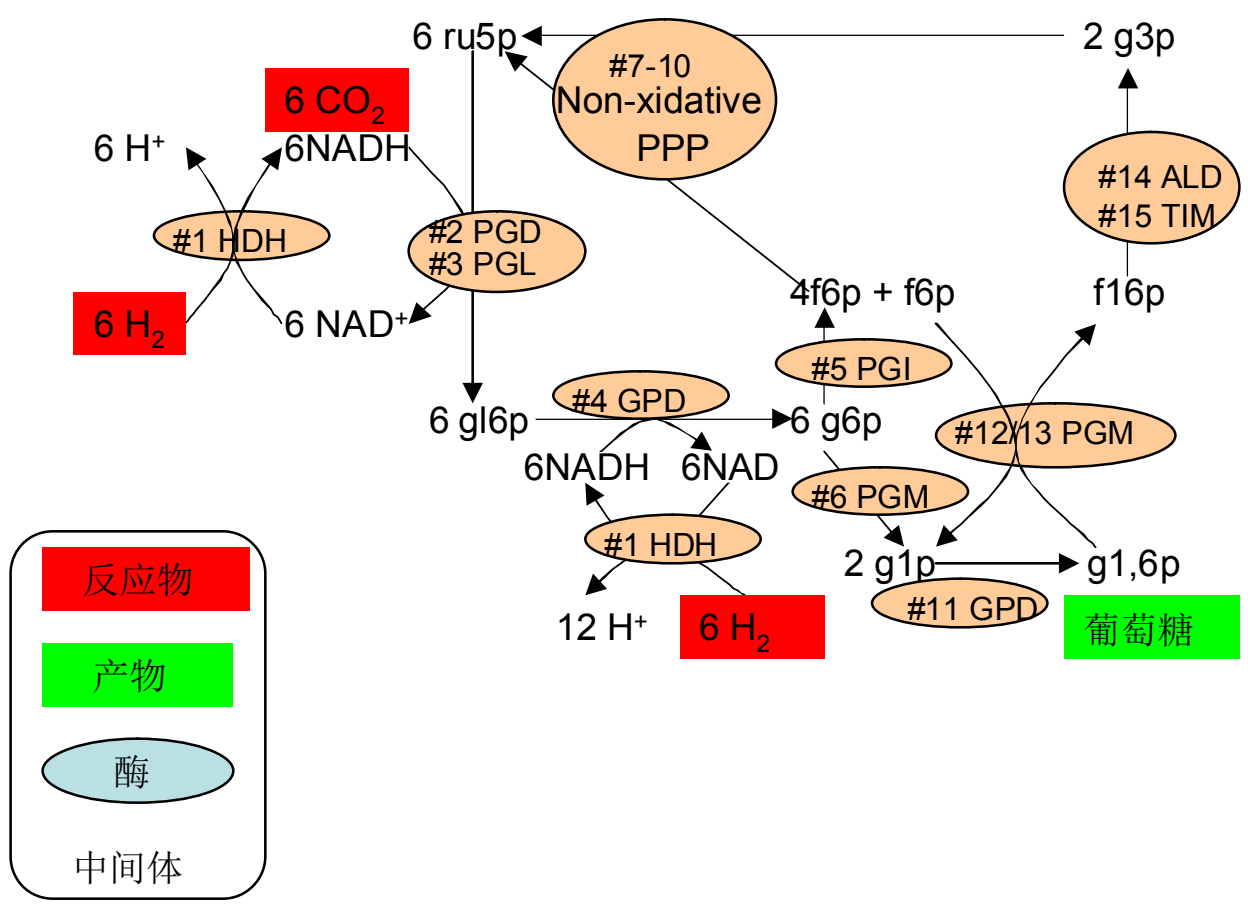

图 4. 电或氢能驱动固定二氧化碳人工光合作用途径：磷酸葡萄糖歧化和固定二氧化碳的磷酸戊糖途径与二磷酸甘油酸缩合途径不同的酶见表 4，中间产物不同的如下：6p g， 6 磷酸葡萄糖酸；gl6p，6 磷酸葡萄糖酸-δ-内酯；省略无机磷酸，$H^+$和水。

在上述三种方案中，我们还可以在磷酸葡萄糖酸脱氢酶催化下，直接将二氧化碳与 5 磷酸戊糖反应，产生 6 磷酸葡萄糖酸(pg6)[22]，6 磷酸葡萄糖酸再经过脱水内酯化[22]和 NADH 还原，生成 6 磷酸葡萄糖[22]，代替上述方案中通过合成甲醛来固定二氧化碳，以及甲醛与磷酸戊糖反应，生成磷酸己糖。主要反应如下：

ru5p + $CO_2$ + NADH → pg6 + $NAD^+$

pg6 → gl6p

gl6p + NADH → hu6p + $NAD^+$

采用二氧化碳直接加成到磷酸戊糖上，可以避免有毒的中间产物甲醛，适合生产作为食品的糖（图 4）。简称固定二氧化碳的磷酸戊糖途径。这个途径只需要使用 3 种酶固定二氧化碳（表 4），而采用甲醛作为中间体固定二氧化碳，需要 4 种酶。

表 4： 电或氢能驱动人工光合作用暗反应途径的反应和使用的酶：固定二氧化碳的磷酸戊糖途径，使用下表反应代替途径 1 中#1—4 反应

| 序号 | 酶编号 | 酶名称 | 反应 | ΔrG'° kJ/m |
|---|---|---|---|---|
| #1 | 1.1.1.44 | 磷酸葡萄糖酸脱氢酶 (PGD) | NAD(P)H + 5 磷酸核酮糖(ru5p) + $CO_2$ → $NAD(P)^+$ + 6 磷酸葡萄糖酸(6pg) | 5.54 |
| #2 | 3.1.1.31 | 6-磷酸葡糖酸内酯酶 (PGL) | 6 磷酸葡萄糖酸(6pg) + $H^+$ → 6-磷酸葡糖酸内酯(gl6p)+ H2O | 22.93 |
| #3 | 1.1.1.49 | 6 磷酸葡萄糖脱氢酶 (GPD) | 6-磷酸葡糖酸内酯(gl6p) + NAD(P)H + $H^+$ → 6-磷酸葡糖(g6p) + $NAD(P)^+$ | 8.37 |

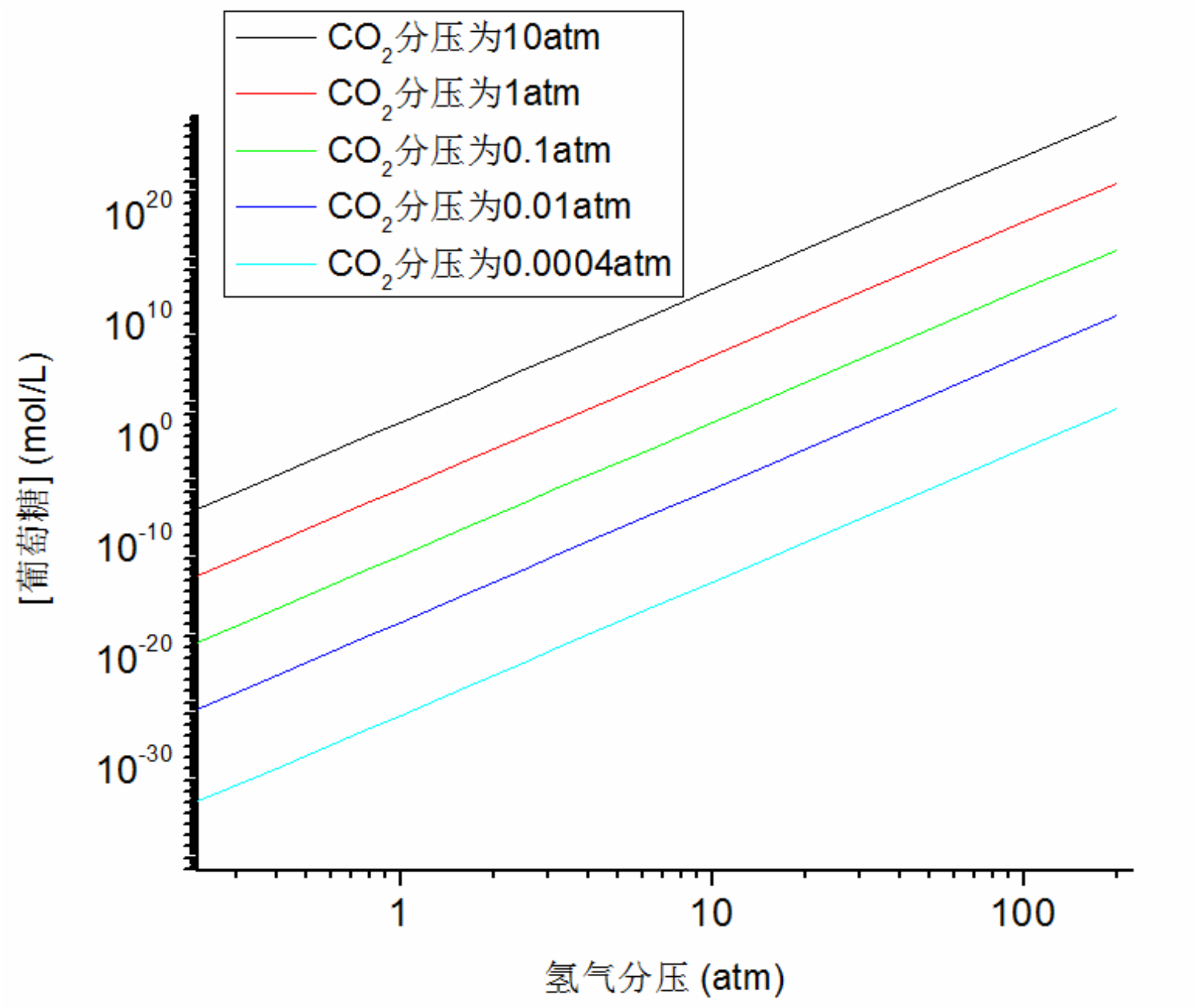

图 5. 氢能驱动人工光合作用暗反应固定二氧化碳生产糖：反应物压力与产物平衡浓度关系

自由能分析：在上述所有方案中，由于反应物均是二氧化碳和氢气（或电解产生的 NADH），产物相同情况下，其标准自由能变化是相同的，产物是淀粉时，均为 44.9kJ/mole 单糖，产物是葡萄糖时，均为 27.kJ/mole 葡萄糖，区别在于中间过程的自由能变化。由于这些中间过程均是自然界存在的生化过程，在不断移除最终产物条件下，均能自发进行。总反应的自由能大于 0，说明标准状态下，平衡是向反应物方向移动。根据反应器是气体，而产物是凝聚态，我们给反应器加压，就能降低总反应自由能，使平衡向产物方向移动。图 5 是不同氢气压力和二氧化碳压力下，人工光合作用反应器内产物平衡浓度。从图中可以看出，产物浓度随反应器压力升高而增加，在 3.49 个大气压下，产物葡萄糖浓度就达到热力学标准状态下 1 mol/L。

三、效率分析

1 热力学效率

在我们提出的人工光合作用暗反应路径中，使用纯氢气和二氧化碳作为反应物，由于总反应自由能大于 0，在常温常压下，产物浓度很低，与热力学标准状态比，需要增加能量，将产物浓度提高到热力学标准浓度 1 mol/L；或者通过输入能量，将氢气分压从 1atm 升高到 2.49 atm，就可以使产物浓度达到热力学标准状态下 1 mol/L 浓度。在这种情况下，系统总压力是 3.49 atm，比标准情况下 2 atm 略高。我们将氢气和二氧化碳送入反应器时，将压力从 2 atm 升高到 3.49 atm，其消耗的能量计入输入能量，用于计算热力学转化效率，得到的人工光合作用暗反应效率是 81.3%。

2 与植物光合作用暗反应热力学效率比较

植物光合作用暗反应是将储存在 NADH 和 ATP 中的自由能转化为糖的化学能。通常储存在 NADH 和 ATP 能量是氧化 NADH 到 $NAD^+$和水解 ATP 到 ADP 后所分别释放的自由能，因此，本文给出的植

物光合作用暗反应的热力学效率是产物糖氧化所能释放的自由能与其反应物 NADH 和 ATP 所能释放的自由能之比，在热力学标准状态下，等于 79.3%。我们提出的人工光合作用暗反应途径，所对应的过程是将 NADH 自由能转化储存在糖中，按照上述根据自由能定义的热力学效率，即 1 摩尔产物糖氧化所能释放的自由能与 12 摩尔 NADH 氧化为 NAD 所释放的自由能之比，计算得到的能量效率是 92.4%。这说明，我们提出的人工光合作用效率比植物光合作用的热力学效率大 13.1%。

在生物体内，ATP 主要来自氧化磷酸化过程，氧化每摩尔 NADH 约产生 2.5 摩尔 ATP[1]，如果以 NADH 作为植物光合作用反应物起点，按照氧化磷酸化计算 ATP 的能量消耗，则植物光合作用暗反应效率仅是 57.7%。实际植物光合作用过程中，再生 3 摩尔 ATP 和再生 2 摩尔 NADH 均需要 4 摩尔光子[1]，相当于每摩尔 NADH 只能再生 1.5 摩尔 ATP，从 NADH 出发，根据光合磷酸化计算光合作用暗反应的热力学效率就更低了。

3 实际人工光合作用暗反应的热力学效率

大气中二氧化碳丰富，但是浓度很低，仅约 390ppm，如果用于人工光合作用，比使用纯二氧化碳，需要消耗更多的能量。但是，高纯度二氧化碳来源很少，而燃烧煤炭，天然气或生物质等的热电厂废气，浓度约为 5－15%，比大气二氧化碳浓度高得多，应是今后人工光合作用生产糖的主要原料。纯化热电厂废气中二氧化碳，将增加能耗和成本，从而降低能量效率和经济性[24-27]。因此，在实际生产中，更可能直接使用未经纯化的燃料集中燃烧产生的废气中的二氧化碳。

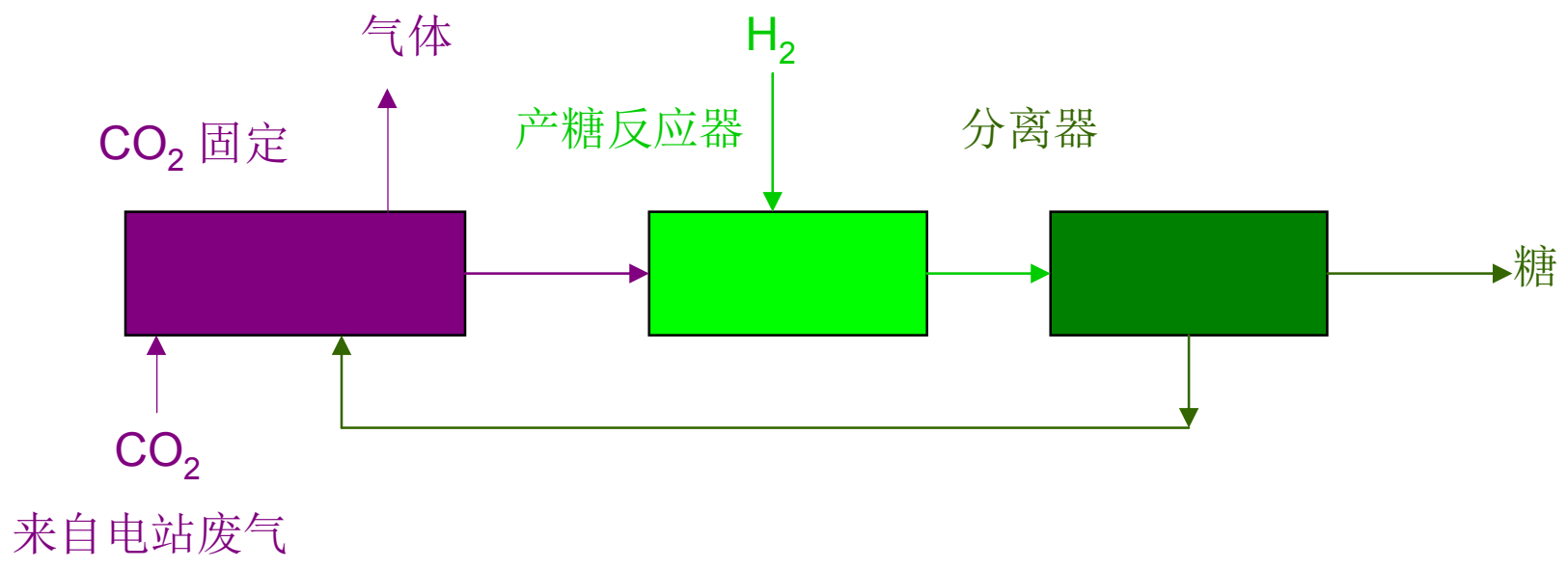

图 6. 氢能驱动人工光合作用暗反应固定二氧化碳生产糖工艺

如图 6 是我们提出的一种使用氢能驱动，实现人工光合作用暗反应固定二氧化碳生产糖的工艺。由于使用热电厂废气作为二氧化碳原料，含有大量其他气体，需要在反应结束后，将其排出，因此，二氧化碳固定反应放在一个独立的反应器进行，即可防止杂质氧气对其他酶的损害，又不会在排放气体时，带走未反应的氢气，从而降低效率。

另设一个反应器进行其他所有反应。使用纯度很高的氢气作为原料，高纯度氢气主要来自电解所得[28]，未来还可来自太阳能转化为热能，在催化剂作用下，热解水产生的高纯度氢气[29-30]。由于使用纯氢气作为原料，它们在反应中都进入液相，这个反应器不需要排出气体，产物随液相排出反应器，经

分离后，又重新回流到固定二氧化碳反应器。由于氢气在水中溶解度很低，从而可以认为，氢气接近完全转化利用。

假设反应器工作温度为 298K，原料中二氧化碳浓度为 10%，二氧化碳利用率为 90%，从反应器排出的废气中，二氧化碳浓度应为 1%，这时，需要将氢气增压到 24.9atm，才能使产物葡萄糖的平衡浓度达到 1mol/L（18%）左右，将氢气增压能耗计入输入能量，得到热力学平衡状态下的效率是 79.6%，比热力学标准状况下效率略低。如果二氧化碳利用率降低到 50%，则反应器排出的废气中二氧化碳浓度为 5%，这时需要将氢气增压到 11atm，热力学效率是 80.2%。

4 人工光合作用总效率

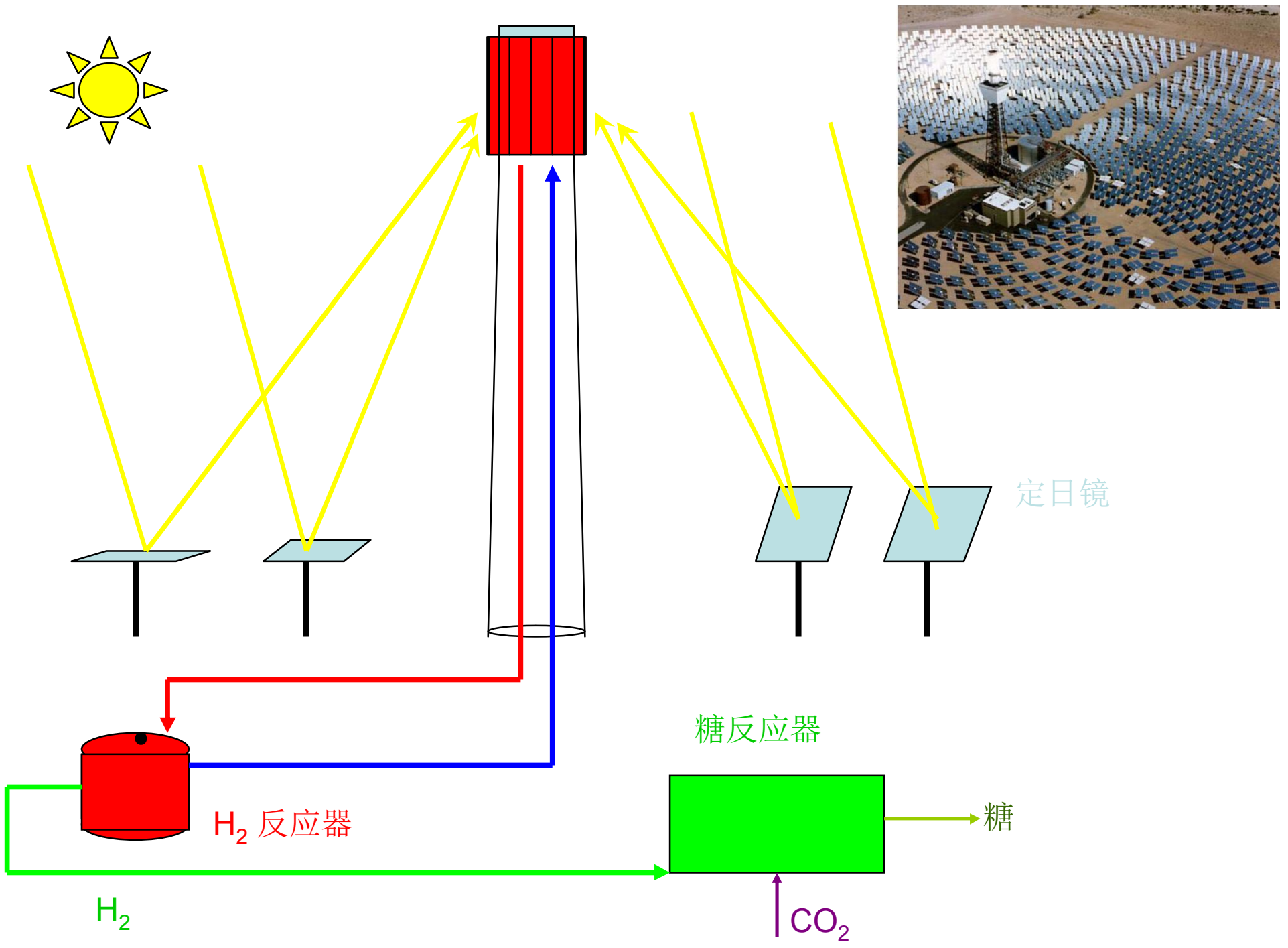

图 7. 太阳能驱动人工光合作用固定二氧化碳生产糖工艺

完整的人工光合作用应从太阳能出发，例如，将太阳能转化为电能或氢能，然后驱动人工光合作用暗反应，得到产物糖。近年来，人们发展了多种太阳能生产电能和氢能方法[31]，如光伏电池直接转换为电能，聚光光伏发电，聚光热发电又分为槽式，蝶式和塔式，电能电解产氢，热能和电能共同驱动，以及热能驱动催化热解产氢，光催化直接分解水产氢等。这里我们以聚光太阳能催化热解产氢，驱动人工光合作用暗反应为例，估算总效率，系统如图 7 所示。

太阳能由多个定日镜反射到中心接收器上，转换为热能，热能驱动热化学过程，分解水产生高纯度氢气，氢气输送到人工光合作用反应器，与来自热电厂废气中二氧化碳一起，在酶催化下，合成糖。这个方案可以规模化，从而可以通过扩大规模降低成本。太阳能到氢能效率可以达到 20%以上[6]，通常胞外酶催化法的效率接近热力学效率[33-34]，从而系统总效率可达到 15%以上。这比植物光合作用效率提高了一个数量级以上。如果采用电解法产氢，效率是 70%[32]，未来多结砷化镓光伏电池的光电转化效率可达到 60%[5]，而胞外酶催化法，副反应少，效率接近热力学效率[33-34]，所以，太阳能驱动人工光合作用总效率最高可达到 30%以上。

四、讨论

胞外酶工程法人工合成糖的主要问题之一是酶，辅酶和中间产物稳定性问题。为了降低它们的生产

能耗和成本在生产产品糖中的比例，除改进生产方法，降低它们的生产成本和能耗外，我们还可以提高它们的稳定性，优化人工光合作用反应条件，增加合成糖反应速率，或/和降低它们的用量，从而提高转化数。转化数每提高 10 倍，催化剂和中间产物对能耗和成本的贡献就会降低 10 倍。转化数达到 100 万以上，酶和中间产物的生产能耗和成本对最终产物糖的能耗和成本的贡献就可以忽略不计[34]。

提高酶的转化数的方法之一是提高酶的稳定性，主要包括使用高温酶，酶固定化等。在 100˚C 以上水中生活的高温菌所产生的酶比通常酶要稳定，例如，从高温菌发展的两种酶，phosphoglucomutase，和 6-phosphogluconate dehydrogenase[35]，在 60 和 80˚C 工作时，转化数均超过 40 万[36]；固定化的葡萄糖异构酶的转化数则达到了 110 万[34]。

ATP 容易水解而减少，我们提出的第三方案，不需要 ATP 参与，第二种方案可以增殖 ATP，补充其消耗，从而可以解决 ATP 水解带来的负面影响。

对磷酸脂类中间体，可以在 maltodextrin phosphorylase（2.4.1.1） 催化下，使用磷酸与淀粉反应，生产磷酸葡萄糖[37]来补充。对于不稳定的磷酸二羟基丙酮，可以采用复合酶和通道技术[34]，及时转化，防止降解。

NADH 容易分解，如何解决 NADH 稳定性问题，是本方法的主要难点之一，可能的解决途径如下：

1． 化学合成甲醛，从而不需要使用 NADH：在我们提出的人工光合作用暗反应途径**磷酸葡萄糖歧化途径**中，只有再生 NADH 及合成甲醛时，需要使用 NADH，其他过程与 NADH 无关（参见图 3 中虚框部分）。如果我们发展了高效化学合成法，将二氧化碳加氢得到甲醛，代替生化转化，就可以避免使用不稳定的辅酶 NADH。

目前从氢气和二氧化碳出发，生成甲醛的方法是先将氢气和二氧化碳合成甲醇[38]，甲醇在部分氧化或脱氢得到甲醛[39]，这个过程的能量转化效率较低。但是，使用二氧化碳和氢气直接合成甲醛，最近也取得了进展，例如，采用双环戊二烯镍为催化剂，以丁醇水混合液为溶剂，甲醛选择性可达到 100%[40]。附载在二氧化硅上的铂铜催化剂催化二氧化碳选择性加氢，在423K和6个大气压下，可得到摩尔比20：87 的甲醇和甲醛混合物，甲醛选择性超过了70%[41]。

2． NADH改性：在生产亮氨酸和苯并氨酸时，使用链接到聚乙二醇上的NADH，转化数可以达到8万和60万[42-44]。在一些有机溶剂中，NAD/NADH也很稳定，有报道，在有机溶剂中进行的有机合成，转化数高达1百万[45]。目前在胞外酶工程法生产手性药物方面，NADH成本已下降到忽略不计的程度[46]。

3． 使用稳定性更好的仿制品替代更换NADH：可以同样克服NAD等易分解，寿命短问题，例如，Lowe等发展了一系列仿制品[47-48]，它们在结果上与NAD相近，容易合成，反应性能相近， 已经能与野生型酶共同协作完成催化作用[49-50]。

## 五 总结

发展人工光合作用，实现工业化粮食生产，摆脱对环境和天气的依赖，是人类梦寐以求的理想。本文根据胞外酶工程法，组合自然界存在的生化过程，提出了 9 种人工光合作用暗反应途径，与太阳能转化生产电能和氢能技术结合，提供了一种替代光合作用生产糖的技术方案。能否实现规模化工业生产，替代农业，为人类供应粮食，还需要今后的实践。我们应尽快通过试验验证，同时解决酶和辅酶稳定性和生产等问题。

此外，利用氢气和二氧化碳化学合成甲醛，利用甲醛合成葡萄糖，仅需要 9 种酶，不需要不稳定性的 NADH 和 ATP 参与，从而容易实现，该技术与胞外酶催化糖产氢技术一起，可使糖成为良好的储能载体。

**致谢：作者的部分工作得到了安徽省发改委支持，特此感谢。**

## Reference

1. Berg, J.M., J.L. Tymoczko, and L. Stryer, *Biochemistry*. 5th ed. 2002, New York: W.H. Freeman. 1 v. (various pagings).
2. *Singhal, G.S., et al., (eds.), Concepts in Photobiology: Photosynthesis and Photomorphogenesis, Springer,*

*Dordrecht*. 1999.
3. Zhu XG, Long SP, and O. DR., *What is the maximum efficiency with which photosynthesis can convert solar energy into biomass?* Curr Opin Biotechnol., 2008. **19**(2): p. 135-9.
4. Pace, R.J., *An Integrated Artificial Photosynthesis Model*, in *Artificial Photosynthesis*, A.F.C.a.C. Critchley, Editor. 2005, WILEY-VCH Verlag GmbH & Co. KGaA: Weinheim.
5. Brown, A.S. and M.A. Green, *Limiting efficiency for current-constrained two-terminal tandem cell stacks.* Progress in Photovoltaics, 2002. **10**(5): p. 299-307.
6. Charvin, P., et al., *Analysis of solar chemical processes for hydrogen production from water splitting thermochemical cycles.* Energy Conversion and Management, 2008. **49**(6): p. 1547-1556.
7. Wendell, D., J. Todd, and C. Montemagno, *Artificial Photosynthesis in Ranaspumin-2 Based Foam DOI: 10.1021/nl100550k.* nano letters, 2010.
8. Zhang, Y.-H.P. and W. Huang, *Electricity-Carbohydrate-Hydrogen (ECHo) Cycle 1 for Sustainability.* Energy & Environmental Science, 2010. **accepted**.
9. Berg, I.A., et al., *Autotrophic carbon fixation in archaea.* Nat Rev Microbiol, 2010. **8**(6): p. 447-60.
10. DiCosimo, R., et al., *Enzyme-catalyzed organic synthesis: electrochemical regeneration of NAD(P)H from NAD(P) using methyl viologen and flavoenzymes.* The Journal of Organic Chemistry, 1981. **46**(22): p. 4622-23.
11. Yoon, S.K., et al., *Laminar flow-based electrochemical microreactor for efficient regeneration of nicotinamide cofactors for biocatalysis.* Journal of the American Chemical Society, 2005. **127**(30): p. 10466-10467.
12. Vuorilehto, K., S. Lutz, and C. Wandrey, *Indirect electrochemical reduction of nicotinamide coenzymes.* Bioelectrochemistry, 2004. **65**(1): p. 1-7.
13. Obert, R. and B.C. Dave, *Enzymatic conversion of carbon dioxide to methanol: Enhanced methanol production in silica sol-gel matrices.* Journal of the American Chemical Society, 1999. **121**(51): p. 12192-12193.
14. Ei-Zahab, B., D. Donnelly, and P. Wang, *Particle-tethered NADH for production of methanol from CO2 catalyzed by coimmobilized enzymes.* Biotechnology and Bioengineering, 2008. **99**(3): p. 508-514.
15. Kuwabata, S., R. Tsuda, and H. Yoneyama, *Electrochemical Conversion of Carbon-Dioxide to Methanol with the Assistance of Formate Dehydrogenase and Methanol Dehydrogenase as Biocatalysts.* Journal of the American Chemical Society, 1994. **116**(12): p. 5437-5443.
16. Kato, N., H. Yurimoto, and R.K. Thauer, *The physiological role of the ribulose monophosphate pathway in bacteria and archaea.* Biosci Biotechnol Biochem, 2006. **70**(1): p. 10-21.
17. Orita, I., et al., *The ribulose monophosphate pathway substitutes for the missing pentose phosphate pathway in the archaeon Thermococcus kodakaraensis.* J Bacteriol, 2006. **188**(13): p. 4698-704.
18. Zhang, Y.-H.P., et al., *High-yield hydrogen production from starch and water by a synthetic enzymatic pathway* PLoS One, 2007. **2**(5): p. e456.
19. Stryer, L., *Biochemistry*. 3rd ed. 1988, New York: W.H. Freeman. xxxii, 1089 p.
20. Schlegel, H.G. and B. Bowien, *Autotrophic bacteria*. Brock/Springer series in contemporary bioscience. 1989, Madison, WI Science Tech Publishers ; Springer-Verlag.
21. Lehninger, A.L., D.L. Nelson, and M.M. Cox, *Lehninger principles of biochemistry*. 4th ed. 2005, New York: W.H. Freeman. 1 v. (various pagings).
22. Caspi, R., et al., *The MetaCyc database of metabolic pathways and enzymes and the BioCyc collection of pathway/genome databases.* Nucleic Acids Research, 2010. **38**: p. D473-D479.
23. Passonneau, J.V., et al., *Glucose 1,6-diphosphate formation by phosphoglucomutase in mammalian tissues.* J Biol Chem, 1969. **244**(3): p. 902-9.
24. Matsumiya, N., et al., *Cost evaluation of CO2 separation from flue gas by membrane-gas absorption hybrid system using a hollow fiber membrane module.* Kagaku Kogaku Ronbunshu, 2005. **31**(5): p. 325-330.
25. Matsumiya, N., et al., *Evaluation of energy consumption for separation of CO2 in flue gas by hollow fiber facilitated transport membrane module with permeation of amine solution.* Separation and Purification Technology, 2005. **46**(1-2): p. 26-32.
26. Aaron, D. and C. Tsouris, *Separation of CO2 from flue gas: A review.* Separation Science and Technology, 2005. **40**(1-3): p. 321-348.
27. Yan, S.P., et al., *Comparative analysis of CO2 separation from flue gas by membrane gas absorption*

*technology and chemical absorption technology in China.* Energy Conversion and Management, 2008. **49**(11): p. 3188-3197.
28. Goswami, D.Y., et al., *A review of hydrogen production technologies.* Fuel Cell Science, Engineering and Technology, 2003: p. 61-74.
29. Steinfeld, A., *Solar thermochemical production of hydrogen - a review.* Solar Energy, 2005. **78**(5): p. 603-615.
30. Abanades, S., et al., *Screening of water-splitting thermochemical cycles potentially attractive for hydrogen production by concentrated solar energy.* Energy, 2006. **31**(14): p. 2805-2822.
31. Goswami, D.Y., *Advances in Solar Energy An Annual Review of Research and Development Vol 17*. 2007, Boulder, CO: American Solar Energy Society.
32. Turner, J.A., *Sustainable hydrogen production.* Science, 2004. **305**(5686): p. 972-974.
33. Y-HP, Z., *Using extremophile enzymes to generate hydrogen for electricity.* Microbe, 2009. **4**: p. 560-5.
34. Zhang, Y.H.P., *Production of Biocommodities and Bioelectricity by Cell-Free Synthetic Enzymatic Pathway Biotransformations: Challenges and Opportunities.* Biotechnology and Bioengineering, 2010. **105**(4): p. 663-677.
35. Wang, Y.R. and Y.H.P. Zhang, *Overexpression and simple purification of the Thermotoga maritima 6-phosphogluconate dehydrogenase in Escherichia coli and its application for NADPH regeneration.* Microbial Cell Factories, 2009. **8**: p. -.
36. Wang, Y. and Y.H.P. Zhang, *A highly active phosphoglucomutase from Clostridium thermocellum: cloning, purification, characterization and enhanced thermostability.* Journal of Applied Microbiology, 2010. **108**(1): p. 39-46.
37. Keseler, I.M., et al., *EcoCyc: A comprehensive view of Escherichia coli biology.* Nucleic Acids Research, 2009. **37**: p. D464-D470.
38. Olah, G.A., A. Goeppert, and G.K.S. Prakash, *Beyond oil and gas : the methanol economy*. 2006, Weinheim an der Bergstrasse, Germany: Wiley-VCH. xiv, 290 p.
39. Reuss, G., et al., *Formaldehyde*. Ullmann's Encyclopedia of Industrial Chemistry. 2002: Wiley-VCH, Weinheim. doi:10.1002/14356007.a11_619.
40. KATSUMI, O. and Y. TADASHI, *PRODUCTION OF FORMALDEHYDE BY CATALYTIC HYDROGENATION OF CARBON DIOXIDE*. 1996.
41. Lee, D.K., D.S. Kim, and S.W. Kim, *Selective formation of formaldehyde from carbon dioxide and hydrogen over PtCu/SiO2.* Applied Organometallic Chemistry, 2001. **15**(2): p. 148-150.
42. W, H., et al., *Isolation of L-phenylalanine dehydrogenase from Rhodococcus sp M4 and its applicationfor the production of L-phenylalanine.* Appl Microbiol Biotechnol, 1987. **26**(5): p. 409-416.
43. Kragl, U., et al., *Enzyme engineering aspects of biocatalysis: Cofactor regeneration as example.* Biotechnology and Bioengineering, 1996. **52**(2): p. 309-319.
44. Wichmann, R. and D. Vasic-Racki, *Cofactor regeneration at the lab scale.* Technology Transfer in Biotechnology: From Lab to Industry to Production, 2005. **92**: p. 225-260.
45. Kazandjian, R. and A. Klibanov, *Regioselective oxidation of phenols catalyzed by polyphenol oxidase in chloroform.* J Am Chem Soc, 1985. **107**: p. 5448-5450.
46. Moore, J.C., et al., *Advances in the enzymatic reduction of ketones.* Accounts of Chemical Research, 2007. **40**(12): p. 1412-1419.
47. Ansell, R.J., D.A.P. Small, and C.R. Lowe, *Synthesis and properties of new coenzyme mimics based on the artificial coenzyme CL4.* Journal of Molecular Recognition, 1999. **12**(1): p. 45-56.
48. Ryan, J.D., R.H. Fish, and D.S. Clark, *Engineering Cytochrome P450 Enzymes for Improved Activity towards Biomimetic 1,4-NADH Cofactors.* Chembiochem, 2008. **9**(16): p. 2579-2582.
49. Lo, H.C. and R.H. Fish, *Biomimetic NAD(+) models for tandem cofactor regeneration, horse liver alcohol dehydrogenase recognition of 1,4-NADH derivatives, and chiral synthesis.* Angew Chem Int Ed Engl, 2002. **41**(3): p. 478-81.
50. Lutz, J., et al., *Bioorganometallic chemistry: biocatalytic oxidation reactions with biomimetic NAD(+)/NADH co-factors and [Cp*Rh(bpy)H](+) for selective organic synthesis.* Journal of Organometallic Chemistry, 2004. **689**(25): p. 4783-4790.
51. Jankowski, M.D., et al., *Group contribution method for thermodynamic analysis of complex metabolic networks.* Biophysical Journal, 2008. **95**(3): p. 1487-99.
52. Leskovac, V., et al., *Thermodynamic properties of the Calvin cycle and pentose phosphate pathway.* Indian

Journal of Biochemistry & Biophysics, 2008. **45**(3): p. 157-165.